# An Investigation of Large Language Models and Their Vulnerabilities in Spam Detection

Qiyao Tang
qtang16@jh.edu
Johns Hopkins University
Baltimore, Maryland, USA

Xiangyang Li[*]
xyli@jhu.edu
Johns Hopkins University
Baltimore, Maryland, USA

**Abstract**

Spam messages continue to present significant challenges to digital users, cluttering inboxes and posing security risks. Traditional spam detection methods, including rules-based, collaborative, and machine learning approaches, struggle to keep up with the rapidly evolving tactics employed by spammers. This project studies new spam detection systems that leverage Large Language Models (LLMs) fine-tuned with spam datasets. More importantly, we want to understand how LLM-based spam detection systems perform under adversarial attacks that purposefully modify spam emails and data poisoning attacks that exploit the differences between the training data and the massages in detection, to which traditional machine learning models are shown to be vulnerable. This experimentation employs two LLM models of GPT2 and BERT and three spam datasets of Enron, LingSpam, and SMSspamCollection for extensive training and testing tasks. The results show that, while they can function as effective spam filters, the LLM models are susceptible to the adversarial and data poisoning attacks. This research provides very useful insights for future applications of LLM models for information security.

## 1 Introduction

Spam messages have become significant challenges to computer users. They not only clutter inboxes but also pose security risks. Rules-based, collaborative, and content-based spam detection approaches have been proposed, but these methods are brittle and hard to generalize. Then spam detections combining machine learning are explored for better accuracy and efficiency.

However, developing an adaptable and accurate spam detection system remains a significant challenge due to the constantly evolving strategies employed by spammers. With the recent development of AI technologies, this problem may be tackled by leveraging Large Language Models (LLMs), trained deep learning models with remarkable capabilities to comprehend and learn from the structural and semantic patterns found in natural language data.

### 1.1 Motivation

LLMs are trained on diverse natural language datasets. Fine-tuning these models on curated spam datasets is essential to enhance their specificity and effectiveness in spam detection. Moreover, not enough efforts have been devoted to understanding the vulnerabilities and robustness of LLM-based spam filters.

This study aims to bridge this gap by fine-tuning LLMs for spam classification and assessing their susceptibility to adversarial attacks and data poisoning attacks. By developing and testing these attack strategies, we evaluate the robustness of LLM-based systems and discusses potential defense mechanisms to mitigate these adversarial threats.

This paper addresses the following three research questions:

(1) Can LLM models support the development of spam detection systems that effectively classify ham and spam emails or messages?
(2) Are LLM-based spam detection models vulnerable to adversarial attacks that modify spam messages with special words or sentences to increase their chance of bypassing detection?
(3) Is an LLM-based spam detection model's performance affected by the training data that has different information characteristics from the emails or messages in detection?

### 1.2 Objectives

Spam detection systems, even those powered by advanced machine learning and AI models, are not immune to adversarial attacks. One particularly effective attack involves the insertion of carefully chosen "good words" or "magic words" that exploit the weaknesses of a model [1][2]. These words are purposefully identified to manipulate the decision boundary of a classifier, causing an error in its prediction. Moreover, differences in the linguistic patterns and contextual features of the emails or messages used to train a spam detection model from those being scrutinized in deployment may obstruct a spam filter from making correct classifications. This disparity is exploited by data poisoning attacks, where maliciously crafted training data is introduced to compromise the model's ability to generalize in detection.

In this project, we build several spam detection systems to examine their effectiveness. Moreover, we explore the impact of the above adversarial and data poisoning attacks on these LLM-based spam detection models.

- This project studies the capabilities of two LLMs of GPT2 and BERT for spam detection. A set of spam filters are trained by fine-tuning these two base LLMs using three spam datasets of Enron, LingSpam, and SMSspamCollection to optimize their classification performance. Then we compare their spam detection performance on these datasets.
- Through a black-box adversarial attack, we apply "magic words" identified for an SVM classifier to attack the LLM spam detection models. These words are found by an approach that employs Projected Gradient Descent (PGD) perturbations to the TF-IDF features of spam emails, as described in [3]. We place these words or composed sentences in different positions within spam emails, e.g., the beginning, after the first sentence, or after the second sentence, etc., to observe their effects on the models' performance. This analysis not only highlights the vulnerabilities inherent in LLMs but also provides insights into the design of more robust spam detection mechanisms capable of resisting adversarial manipulation.
- We simulate data poisoning attacks by training a spam detection model on one dataset, such as LingSpam, but testing it against heterogeneous emails or messages, such as those from a different dataset of Enron. This cross-dataset evaluation aims to examine the effect of differences in training and testing data on model performance in a scenario similar to data poisoning. By analyzing the model's misclassifications, we seek to understand the vulnerabilities that arise from dataset-specific biases. Such insights can be useful to develop strategies to mitigate the related risks.

## 2 Related Work

Early spam detection approaches relied heavily on rules-based systems and statistical models which, often struggled with adaptability to new spam tactics. The introduction of machine learning brought methods offered improved generalization but were limited in handling the complexity of modern spam content.

Recent advancements in deep learning, particularly the emergence of LLMs, have significantly enhanced the ability to understand the semantic and syntactic structures of language. These models have shown promise in spam detection tasks by leveraging pre-trained knowledge and fine-tuning on domain-specific datasets. However, the robustness of LLM-based spam detection systems against adversarial attacks and their generalizability to spam data remain understudied areas of research.

### 2.1 Spam Detection Using Machine Learning

Leading email service providers, including Gmail, Yahoo Mail, and Outlook have long relied on advanced machine learning (ML) techniques, to enhance their spam filtering systems. For instance, Google's spam detection models, which incorporate tools like Google Safe Browsing, achieve a detection rate of 99.9%, allowing only one in a thousand spam messages to bypass their filters [4].

These models leverage vast amounts of data to identify and classify spam and phishing emails with remarkable accuracy. Unlike traditional rule-based systems, modern ML-based spam filters dynamically generate new detection rules based on continuous learning from incoming messages, employing several approaches:

- *Content-based Filtering:* Content-based filtering is a widely employed method for automatically generating filtering rules and classifying emails using various machine learning algorithms, such as Naïve Bayes and Support Vector Machines (SVM). This approach typically examines the text within emails, analyzing the frequency, distribution, and patterns of words and phrases. The insights derived from this analysis are then utilized to create rules that help in identifying and blocking spam emails effectively [5].
- *Case-based/Sample-based Filtering:* This widely used approach begins by collecting all emails, including spam and non-spam. The processing transforms emails, extracts and selects relevant features, and groups the data into two vector sets. Finally, machine learning algorithms are employed to train and test the datasets, enabling the classification of incoming emails as either spam or non-spam [5].
- *Rule-based Filtering:* This method relies on predefined rules to analyze patterns within emails, often employing regular expressions to match specific message characteristics. When a message aligns with multiple patterns, its score increases, whereas mismatched patterns lead to a reduction in the score. Messages with scores exceeding a certain threshold are classified as spam. While some rules remain static, others require frequent updates to adapt to evolving spam techniques, as spammers continuously devise new strategies to bypass email filters [4].

### 2.2 Spam Detection Using Large Language Model

By understanding and learning complex semantic and structural patterns in text, recent advancements in the use of Large Language Models (LLMs) have demonstrated significant potential in improving email spam detection. In a recent study [6], the authors compare the performance of LLMs such as RoBERTa, SetFit, and Flan-T5.

- *RoBERTa:* A refined version of BERT, RoBERTa improves training stability and performance by leveraging larger datasets and longer training times. It has been used to classify email text with high precision.
- *SetFit:* SetFit (Sentence Transformer Fine Tuning) employs contrastive learning to create high-quality embeddings with fewer parameters. Its lightweight nature allows efficient deployment while achieving competitive accuracy.
- *Spam-T5:* Spam-T5, a fine-tuned version of Flan-T5, has shown state-of-the-art performance in spam detection tasks. By framing the task as a sequence-to-sequence problem, Spam-T5 generates labels like "spam" or "ham" directly, with remarkable results.

This study compared LLMs against traditional methods (e.g., Naïve Bayes, SVM, XGBoost) across datasets like LingSpam, SMSspamCollection, SpamAssassin, and Enron. The results showed that LLMs consistently outperformed traditional machine learning methods. Specially, Spam-T5 achieved the highest levels of accuracy, performing well not only when trained with large datasets but also excelling in few-shot learning scenarios where few labeled examples were available. This is particularly important because spam detection systems need to be adaptable and robust, able to identify new types of spam. Additionally, it seems that LLMs are more robust against adversarial attacks, where spammers deliberately alter email content to evade detection by spam filters.

### 2.3 Adversarial Attacks on Spam Detection

Spammers have developed sophisticated techniques in the hope to bypass detection by several distinct strategies [7]:

- *Tokenization:* Spammers disrupt the feature selection process by altering the structure of messages. Common tactics include splitting key words with spaces or leveraging HTML, JavaScript, or CSS layout tricks to obscure the message's true content.
- *Obfuscation:* This involves concealing the message's contents through encoding or misdirection. Techniques include HTML entity encoding, URL encoding, letter substitution, and formats like Base64 or Quoted-printable encoding.
- *Statistical Attacks:* These attacks manipulate message statistics to confuse filters. For example, *Weak Statistical Attacks* Use purely random data, such as inserting random words, fake HTML tags, or nonsensical text excerpts, to distort the message's features. On the other hand, *Strong Statistical Attacks* employ more targeted and informed data to improve success rates. These attacks often rely on feedback mechanisms to refine their methods based on which spam messages bypass filters successfully.

One important study is the "good word attack" [1]. Researchers found that a relatively small number of well-chosen words can be added to a spam email to trick filters, such as those using naive Bayes and maximum entropy models, into classifying it as legitimate. These words are identified by statistically evaluating the effect of different combinations of words on these models.

In another study of "Magic Word attack" [3], the authors proposed a new method of crafting adversarial examples that translates the findings of adversarial manipulations in the feature space, e.g., TF-IDF or Word2Vec vectors, back to changes to be made in the problem space, e.g., emails. Projected Gradient Descent (PGD) can iteratively apply small, constrained perturbations to the features to induce the misclassification by a spam detection model [8]. Then, a small set of "magic words" are determined by examining their corresponding features that, after small perturbations during this process, have the most significant influences on the classification outcomes. These words can be added to spam emails without changing their nature.

In a comprehensive evaluation of this method [2], the researchers utilized different datasets, feature extraction methods, and machine learning models. The technical workflow is shown in Figure 1. The feature extraction methods include TF-IDF, Word2Vec, and Doc2Vec. The PGD perturbations are conducted on a trained SVM model to identify the magic words. In white-box attacks, these words are tested for their effectiveness on the same SVM as the target. In black-box attacks, the target models being attacked include Decision Tree, Logistic Regression, MLP, and an ensemble classifier using these three models. In gray-box attacks, the target model may be a SVM model but using a different feature extraction method. In the last two cases, the adversary has no or limited knowledge of the target spam filter system. The "magic word attack" has proven to be effective on traditional machine learning models in all the attack scenarios.

## 3 Methodology
### 3.1 Large Language Model for Spam Detection
The code used is based on a previous work [9]. The tasks include training and testing LLM models on spam datasets.

**Datasets:** The dataset comprises labeled data of spam emails or short messages, where each entry is categorized as either *spam* or *ham* (non-spam). Only the message content, plus the subject for an email, is used.

**Data Preprocessing:** A tokenizer provided by the specific LLM framework, i.e., Hugging Face's transformers, is used. The tokenizer converts a message to numerical representations in PyTorch-compatible tensor formats.

**Data Splitting and Loader Initialization:** The preprocessed information is divided into training and testing sets using an 80-20 split. The training subset is used to optimize the model, while the testing subset evaluates its performance.

**Model Selection:** The base LLMs are fine-tuned for binary classification tasks by adding a *classification head*, which is a fully connected layer customized for two-class prediction output (spam or ham).

**Evaluation:** At the end of each epoch, the model is evaluated on the testing set to assess its performance. Several performance metrics are collected including *accuracy*, *precision*, *F1 Score*, *false positive rate (FPR)*, and *False Negative Rate (FNR)*. Among them, FNR quantifies the proportion of the spam emails incorrectly classified as ham. This measure is important in the result analysis of adversarial attacks.

### 3.2 Adversarial Attack Using "Magic Words" and "Magic Sentences"

The process of identifying "magic words" follows the steps in Figure 1 for the previously described works [2][3]. A code implementation of this attack is available [10]. Moreover, the specially selected words are made into sentences in the evaluation. We are interested to characterize their effect on LLM-enabled spam detection models that are supposed to understand nuanced linguistic structures.

**Data Processing:** A spam dataset is divided into three subsets: training (64%), validation (16%), and testing (20%). The validation set is needed for identifying the "magic words". The preprocessing step prepares the emails by removing numbers, punctuation, and common English stop words, and applying stemming to reduce words to their root forms.

**Training a Classifier:** This study used the *Term Frequency-Inverse Document Frequency (TF-IDF)* method that represents one email as a vector of TF-IDF values corresponding to all the words in the vocabulary used by a dataset. Then, a *Support Vector Machine (SVM)* classifier is trained using the training subset.

**Finding Magic Words:** The Projected Gradient Descent (PGD) method is used to perturb the TF-IDF feature vectors of a number of randomly selected spam emails from the validation subset. The controlled modifications to these feature vectors may cause the SVM classifier to flip the classification from spam to ham successfully.

Then, the set of "magic words" is the intersection of two sets of words. *Top Perturbation Words* are the words whose features are changed the most during the successful perturbations that flip the classification of a spam email to "ham". *Unique Ham Words* are those words that exclusively appear in ham emails. Intuitively, the features of these words in their intersection increase from zero, as they are not in the original spam emails, to positive values after perturbation, as they appear in the modified feature vectors.

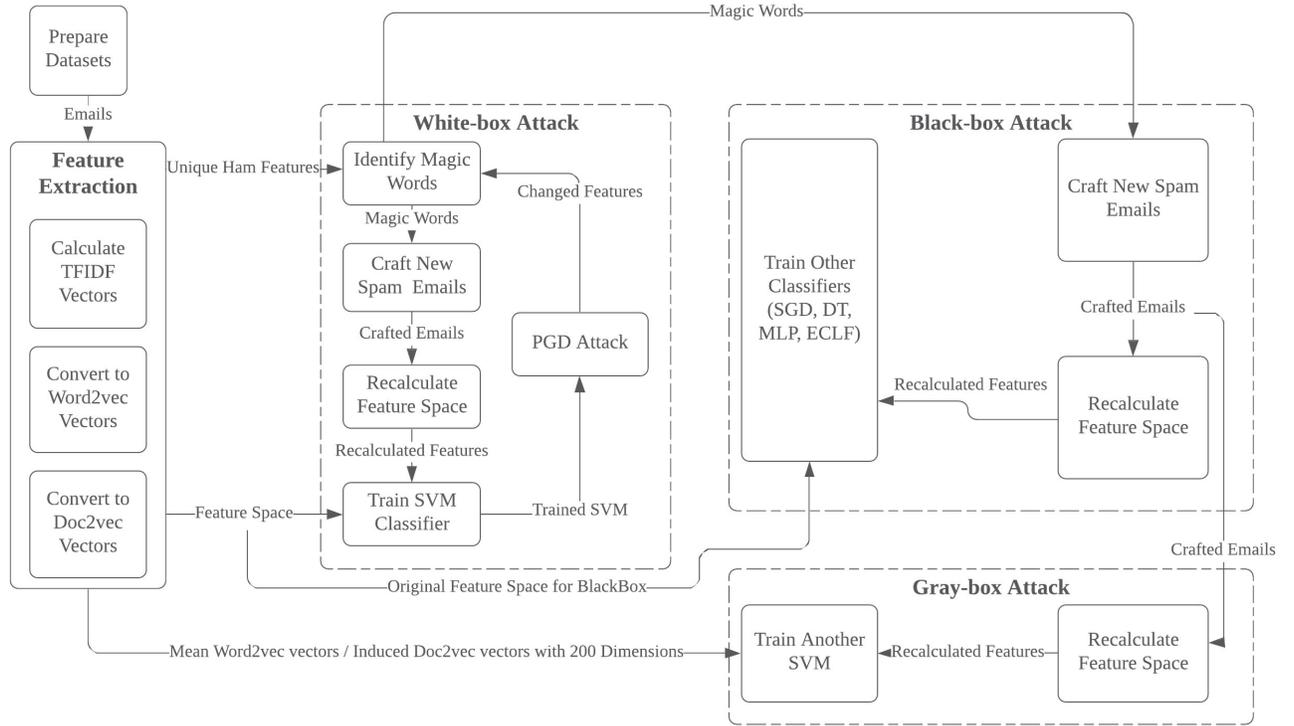

**Figure 1: Adversarial attacks that perturb email features, craft adversarial spam emails by adding specially identified magic words, and conduct white/gray/black-box attacks, as in [2] (This paper studies black-box attacks on LLM-enabled spam classifiers.)**

We can control the number of magic words by limiting the size of *Top Perturbation Words* in consideration. Usually adding a small set of these words has shown to be effective to make a spam email to bypass detection, i.e., resulting in a false negative error. More information on identifying magic words is in [2][3].

**Making "Magic Words" into "Magic Sentences":** The machine learning models evaluated in the previous works consider all the words of an email at once in a rather "coarse" view, without looking at the order of them. So, the structure of combining the magic words when we add them to a spam message does not matter for these spam filters. On the other hand, LLMs perform semantic analysis and comprehend the words in their local context, i.e., the order in which they are associated with each other. As a result, we hypothesize that inserting magic words directly may not be as effective against LLMs as using meaningful and coherent sentences containing these words.

**Injecting Point of "Magic Words" and "Magic Sentences":** Moreover, we examine such attacks systematically by placing these words or sentences at different positions within the message, e.g., at the beginning, after the first, second, and third sentence, or at the end. This is important as LLM-based systems, similar to a human reader, use information seen in the past to process the word embeddings in current attention.

The use of words or sentences and the injection point bear consequence on the overall success of a spam message. Spam detection has two layers. The human email user still processes it even if the message can evade the spam filter. Reading these words simply thrown together, unrelated to the rest of the email and hardly meaningful, likely raises a red flag to the user. At the same time, burying them deep rather than putting them early in the email helps to make them less visible to the reader.

The evaluation adds the magic words or sentences to the spam messages of the same testing subset in Section 3.1, and feeds them to the LLM spam filters, in a black-box attack. For adversarial attacks, an increase in FNR indicates the success of such attacks, as these modified messages bypass the detection with higher probabilities.

### 3.3 Cross-dataset "Data Poisoning" Attack

We train the model on one dataset and test it on another. This approach simulates real-world scenarios where spam filters are exposed to emails in detection that differ from their training data in terms of structure, language patterns, and feature distribution. A significant drop in performance may indicate the model's vulnerability to challenges similar to data poisoning attacks, especially in a dynamic working environment where the nature of spam emails constantly evolves.

## 4 Experimental Design

### 4.1 Use of Datasets

We selected three widely used datasets: *Enron*, *LingSpam*, and *SMSspamCollection*. Each dataset represents different characteristics of email or message communication, providing a diverse testing ground for the experiments.

**Enron Dataset:** The Enron dataset is a large-scale collection of emails. It contains over 33,092 emails, though smaller subsets are often used for spam detection tasks. The portion of the spam and ham is approximately 50% to 50%. Enron dataset is one of the most realistic benchmarks in a business setting for evaluating email-based spam detection models.

**LingSpam Dataset:** The LingSpam dataset is a collection of 2,827 emails, primarily sourced from the Linguist mailing list. 468 of these emails are spam, while the rest are ham emails. This dataset is notable for its focus on professional communication and academic discussions.

**SMSspamCollection:** The SMSspamCollection dataset comprises 5,572 text messages, with approximately 13.4% labeled as spam and the rest as ham. Unlike email datasets, this collection captures the characteristics of mobile communication, including short message length, informal language, and abbreviations. Its focus on SMS communication makes it a resource for evaluating the adaptability of spam detection techniques across different mediums.

### 4.2 Use of Large Language Models

The experiments focused on two LLMs: GPT2 and BERT. The setup involves data preprocessing, model training, validation, and evaluation on multiple metrics.

**Data Preprocessing:** Preprocessing was performed using tokenizers specific to each LLM. For GPT2, the *GPT2Tokenizer* was used, and for BERT, the *BertTokenizer* was employed. Each message was tokenized into input IDs and attention masks with a maximum sequence length of 32 tokens. Special tokens, e.g., *[CLS]*, *[SEP]* for BERT, and *</endoftext/>* for GPT2, were added automatically during tokenization. Tokenized data was split into training and testing sets, and PyTorch TensorDataset and DataLoader were used to create batches for efficient processing.

**Model Initialization:** Two pre-trained models, *BertForSequenceClassification* and *GPT2ForSequenceClassification*, were fine-tuned for binary classification tasks. Both models were initialized with their pre-trained weights and modified to include a classification head for predicting spam and ham labels.

**Training:** Each model was fine-tuned separately for two epochs. During each epoch, the training loss was calculated for every batch using the cross-entropy loss function. The training process used the *AdamW* optimizer with a learning rate of $5 \times 10^{-5}$. The optimizer updated model weights through gradient backpropagation. After each epoch, the models' performance was evaluated on the testing set to monitor progress and prevent data over-fitting.

### 4.3 Experimentation Environment

All experiments were conducted using Python with the *transformers* libraries in the environment of Anaconda 3 Jupyter Notebook. The computational setup included a CPU-enabled system for efficient training and inference. The models were also trained on a GPU-enabled system whenever possible to accelerate computations. Both models shared identical experimental conditions to ensure a fair comparison of their performance.

## 5 Result Analysis

### 5.1 Spam Detection Result

Table 1 summarizes the performance of GPT2 and BERT evaluated across the three datasets of Enron, LingSpam, and SMSspamCollection.

**Enron Dataset:** GPT2 demonstrates consistently low FNR and FPR values, with significant improvements from Epoch 1 to Epoch 2. The Accuracy improved from 98.43% to 98.82%, and the F1 Score also showed a slight increase. BERT Shows higher FNR compared to GPT2 after Epoch 1, but a significant improvement is observed in Epoch 2. Overall, BERT has higher FNR but lower FPR than GPT2, although the differences are small. Such differences may be factors in selecting these models for spam detection, informing a decision maker who puts more weight on detecting all the spams or avoiding missing legitimate messages.

**LingSpam Dataset:** GPT2 struggles in Epoch 1 with a high FNR (11.32%), but Epoch 2 demonstrated notable improvements in all metrics. The F1 Score increases from 89.62% to 91.99%. BERT achieves perfect FNR (0%) in Epoch 1, albeit with a higher FPR (5.19%). However, after Epoch 2, FPR reduces to 0%, resulting in an F1 Score of 98.93%. This makes BERT exceptionally effective on this dataset after sufficient training.

**SMSspamCollection Dataset:** GPT2 achieves high Accuracy in both epochs, with Epoch 2 performing better (99.29%). However, Precision shows a slight trade-off, suggesting sensitivity to the dataset's characteristics. BERT achieves perfect Precision (100%) in Epoch 1 but showed a drop in Epoch 2, accompanied by an increase in FPR and FNR. These variations highlight the model's sensitivity to nuances in this dataset as these text messages contain considerably less information.

The results seem to tell a few useful points about the use of LLMs in spam detection:

- *LLM Model Comparison:* GPT2 generally exhibits robust performance across all datasets, but its FNR is slightly less consistent compared to BERT. Overall, BERT excels in FNR across the dataset, making it a strong candidate if the goal is to block all spams.
- *Dataset Comparison*: The overall trends of these two models seem compatible with each other over the two spam email datasets. However, they let more spam messages through without being flagged for SMSspamCollection, suggesting a challenge likely due to less information in a short text message.
- *Impact of Training Epoch:* In general, both models show notable improvements from Epoch 1 to Epoch 2 across all datasets, with reduced FNR, FPR, and Train Loss values.

### 5.2 Adversarial Attack Result

**Magic Words and Sentences Used:** For the Enron dataset, 11 "magic words" were identified:

**Table 1: Spam Detection Performance Using GPT2 and BERT**

| Dataset | Model | Epoch | FNR | FPR | Accuracy | Precision | F1 Score | Train Loss |
|---|---|---|---|---|---|---|---|---|
| Enron | GPT2 | 1 | 1.15% | 1.99% | 98.43% | 98.09% | 98.45% | 11.18% |
| Enron | GPT2 | 2 | 0.91% | 1.45% | 98.82% | 98.55% | 98.60% | 3.04% |
| Enron | BERT | 1 | 3.51% | 0.60% | 97.95% | 99.38% | 97.99% | 7.83% |
| Enron | BERT | 2 | 1.21% | 0.96% | 98.91% | 99.03% | 98.68% | 1.99% |
| LingSpam | GPT2 | 1 | 11.32% | 0.00% | 97.92% | 94.44% | 89.62% | 15.67% |
| LingSpam | GPT2 | 2 | 3.31% | 0.43% | 98.96% | 93.52% | 91.99% | 2.90% |
| LingSpam | BERT | 1 | 0.00% | 5.19% | 95.77% | 81.54% | 89.20% | 11.82% |
| LingSpam | BERT | 2 | 1.89% | 0.00% | 99.65% | 100.00% | 98.93% | 4.29% |
| SMSspamCollection | GPT2 | 1 | 6.74% | 0.00% | 98.93% | 97.14% | 94.27% | 15.07% |
| SMSspamCollection | GPT2 | 2 | 3.37% | 0.21% | 99.29% | 96.19% | 95.30% | 4.51% |
| SMSspamCollection | BERT | 1 | 6.74% | 0.00% | 98.92% | 100.00% | 96.30% | 8.23% |
| SMSspamCollection | BERT | 2 | 2.25% | 2.13% | 97.85% | 89.69% | 93.42% | 2.49% |

> *sitara cera kaminski kal listbot ena erisk reactionsnet enrononline lavo lokay*

We created the following "magic sentence" from these words:

> *Sitara, Cera, and Kaminski collaborated on a project that utilized ListBot and Ena, analyzing data from Erisk, ReactionsNet, and EnronOnline, while drawing inspiration from Lavo and Lokay.*

The 23 "magic words" identified from the LingSpam dataset are:

> *translation cascadilla workshop proceeding benjamin academic ldc chorus native colingacl french sentence pkzip euralex linguistic risked ammondt phonetic arizona grammar ipa theory linguist*

We crafted the following two sentences based on these words:

> *Academic linguist Benjamin pkzips phonetic sentence translation grammar theory in Euralex COLING/ACL workshop proceeding in Arizona. Native French Ammondt risked the linguistic IPA Chorus of LDC Cascadilla.*

The SMSspamCollection dataset was also analyzed, yielding 69 identified Magic Words. However, this dataset is unique with short text messages. Incorporating a large number of "magic words" significantly alters the structure and nature of a message. Therefore, we did not try attacks on this SMSspamCollection dataset.

**Results:** Our analysis focuses primarily on False Negative Rate (FNR), as it serves as the success rate of these attacks. The results are given in Tables 2 and 3. They are also plotted in Figures 2-5 for easier comparison.

Here we first explain the notations used in the tables and figures.

- *word/sentence@number* indicates the attack method and the injection point.
- *word/sentence* is one of two attack methods, i.e., "magic words" or "magic sentences".
- *@number* is the position of these words/sentences being added to an email, i.e., 0-at the beginning, 1-after the first sentence, 2-after the second sentence, etc.
- *@end* means the very end of the email.
- *None* is for the result under no attack, i.e., baseline FNR.

**Analysis:** Over the ste of experiments, we can clearly see that success rate varies significantly depending on the attack method and the injection point .

- *Effectiveness of Injection Point:* When the "magic words" or "magic sentences" are inserted at the beginning of the email, the FNR consistently reached 100% for all the models trained on both datasets. This indicates a complete failure of the models to flag any spam emails. When these words or sentences are injected later in the emails, e.g., moving to @1, @2, @3, and @end, the attack success rate shows a gradual decline. But it remains elevated compared to the baseline case. For instance, when magic words are injected after first sentence, the FNR ranges between 27% to 53% approximately. However, compared to the original baseline performance, injecting them at the end of the messages only incurs a minor increase in FNR. This positional difference is due to how LLM models use established context information, i.e., seen early on in an email, to impact the later content of the email.
- *Comparison of Word-based and Sentence-Based Attacks:* Interestingly, these two attack methods show different characteristics. Arguably sentence-based attacks are slightly more effective overall than word-based attacks. They maintain a higher impact when the injection point is toward the start of the email, e.g., after the first sentence. However, their impact may drop more quickly in several experiments when the injection point moves down the email. The effectiveness of these two methods is mixed when the injection point is at the end of an email, when it is "too late" to maximize their effect on the processing of this email.
- *LLM Model Comparison:* It is hard to conclude the differences between these two models facing adversarial attacks based on the numbers in the two tables. Examining the plots in the four figures, BERT seems in general has lower FNR rates, i.e., higher resistance to an attack, at intermediate injection points. These injection points, not at the immediate start of an email, are likely more practical choices for spammers in these attacks. How these two models perform also vary for the two datasets. It is certainly worth further investigation of how these and other LLM models respond to these attacks.
- *Impact of Training Epoch*: Training the models for an additional epoch (Epoch 2) leads to mixed results. In fact, for several injection points, the FNR increases after Epoch 2, suggesting that additional training might make the models more susceptible to the adversarial modifications rather than improving their robustness. This shows again that the classification accuracy and resistance to adversarial attacks are two different goals that needs to be managed in a more comprehensive manner.

**Table 2: Adversarial Attack Success Rate on the Enron Dataset**

| Attack Method/Injection Point | GPT2-Epoch1 | GPT2-Epoch2 | BERT-Epoch1 | BERT-Epoch2 |
|---|---|---|---|---|
| word@0 | 100.00% | 100.00% | 100.00% | 100.00% |
| word@1 | 27.31% | 32.84% | 37.95% | 34.50% |
| word@2 | 24.33% | 24.96% | 11.40% | 16.93% |
| word@3 | 8.75% | 10.20% | 5.11% | 13.08% |
| word@end | 1.47% | 5.38% | 3.64% | 2.77% |
| sentence@0 | 100.00% | 100.00% | 100.00% | 100.00% |
| sentence@1 | 44.48% | 49.80% | 52.93% | 37.65% |
| sentence@2 | 17.35% | 23.43% | 26.05% | 23.19% |
| sentence@3 | 10.41% | 11.34% | 4.03% | 4.99% |
| sentence@end | 4.81% | 2.05% | 2.14% | 3.58% |
| None | 1.15% | 0.91% | 3.51% | 1.21% |

**Table 3: Adversarial Attack Success Rate on the LingSpam Dataset**

| Attack Method/Injection Point | GPT2-Epoch1 | GPT2-Epoch2 | BERT-Epoch1 | BERT-Epoch2 |
|---|---|---|---|---|
| word@0 | 100.00% | 100.00% | 100.00% | 100.00% |
| word@1 | 46.94% | 42.86% | 30.61% | 44.90% |
| word@2 | 37.76% | 32.65% | 26.53% | 22.45% |
| word@3 | 18.37% | 20.41% | 20.41% | 12.24% |
| word@end | 8.16% | 10.20% | 10.20% | 4.08% |
| sentence@0 | 100.00% | 100.00% | 100.00% | 100.00% |
| sentence@1 | 48.98% | 52.04% | 50.00% | 44.90% |
| sentence@2 | 27.84% | 26.80% | 22.68% | 10.31% |
| sentence@3 | 15.46% | 18.56% | 20.62% | 20.62% |
| sentence@end | 12.24% | 4.08% | 4.08% | 6.12% |
| None | 11.32% | 3.31% | 0.00% | 1.89% |

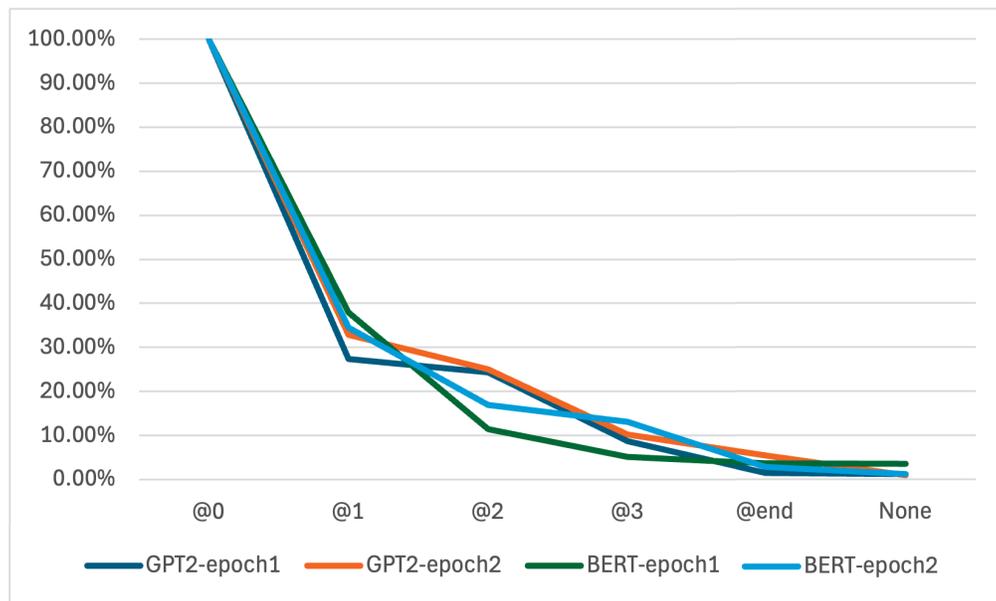

**Figure 2: Success Rate of "Magic Word" Attacks on the Enron Dataset**

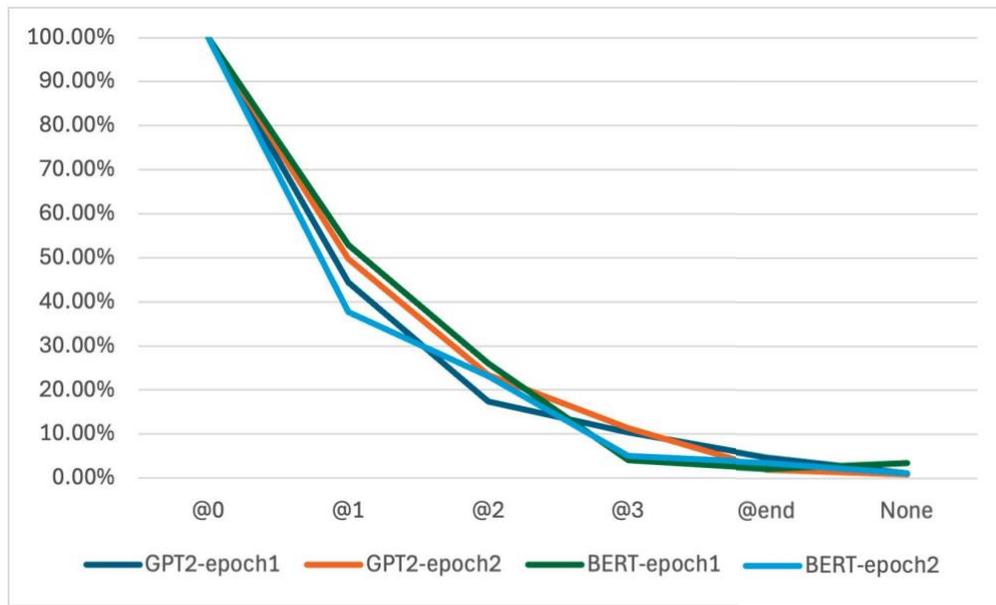

**Figure 3: Success Rate of "Magic Sentence" Attacks on the Enron Dataset**

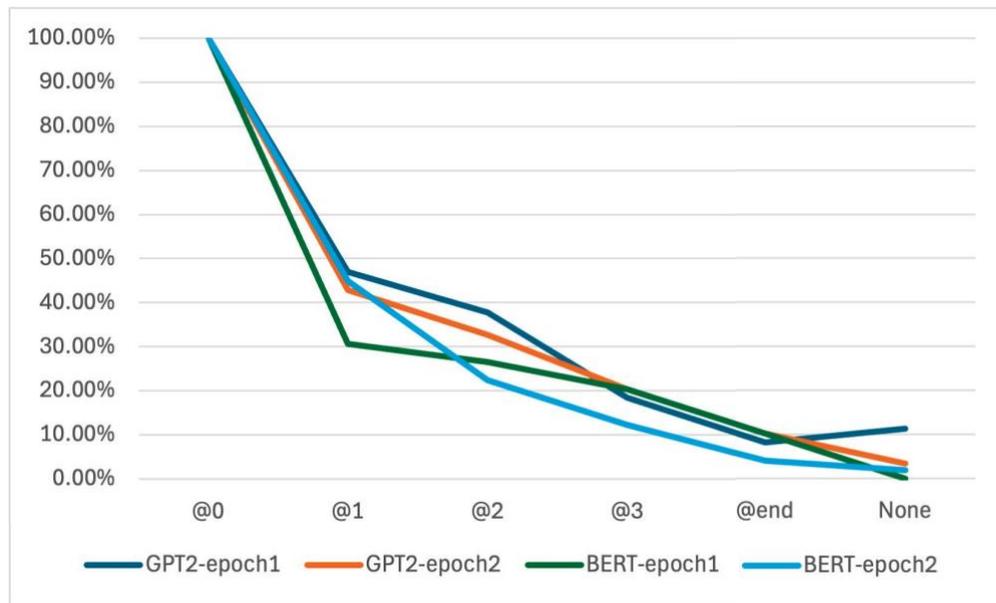

**Figure 4: Success Rate of "Magic Word" Attacks on the LingSpam Dataset**

### 5.3 Cross-dataset Poisoning Attack Result

**Result Analysis:** The results are shown in Tables 4 and 5. The performance of both models drop significantly in these experiments.

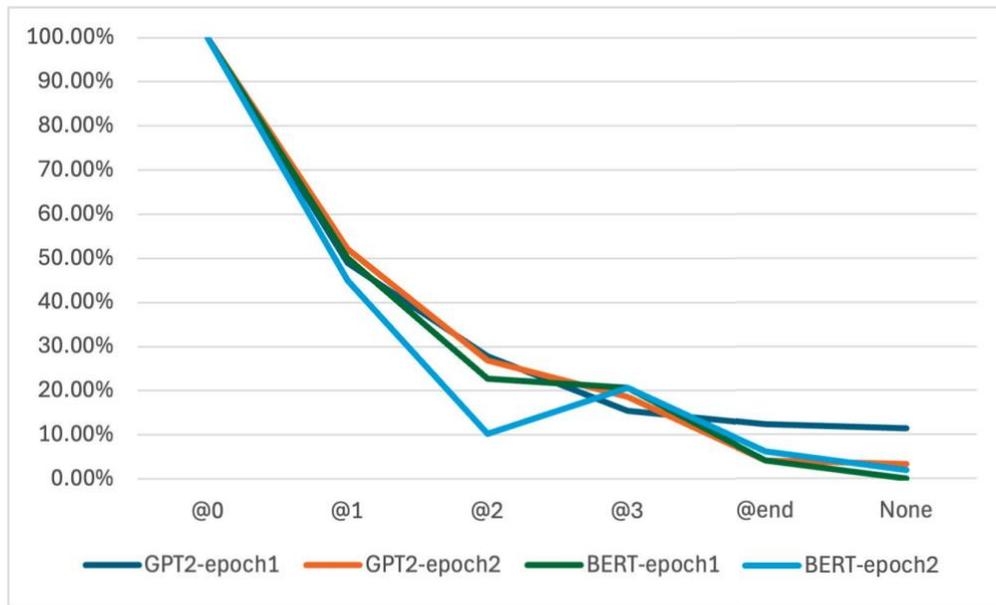

**Figure 5: Success Rate of "Magic Sentence" Attacks on the LingSpam Dataset**

**Table 4: Result of Cross-dataset Poisoning Attacks on GPT2**

| Training Set | Testing Set | Epoch | FNR | FPR | Accuracy | Precision | F1 Score | Train Loss |
|---|---|---|---|---|---|---|---|---|
| LingSpam | Enron | 1 | 26.83% | 26.90% | 73.14% | 73.28% | 73.23% | 30.47% |
| | | 2 | 32.21% | 11.13% | 78.28% | 86.00% | 75.82% | 4.32% |
| Enron | LingSpam | 1 | 9.18% | 15.14% | 85.89% | 55.63% | 68.99% | 11.18% |
| | | 2 | 6.12% | 27.51% | 76.19% | 41.63% | 57.68% | 3.18% |
| LingSpam | SMSspamCollection | 1 | 15.44% | 50.52% | 54.17% | 20.52% | 33.03% | 16.45% |
| | | 2 | 9.40% | 32.19% | 70.85% | 30.27% | 45.38% | 2.94% |
| SMSspamCollection | LingSpam | 1 | 68.37% | 14.71% | 76.01% | 31.00% | 31.31% | 14.58% |
| | | 2 | 67.35% | 11.51% | 78.84% | 37.21% | 34.78% | 3.84% |
| Enron | SMSspamCollection | 1 | 22.15% | 23.29% | 76.86% | 34.02% | 47.35% | 17.06% |
| | | 2 | 2.01% | 71.22% | 38.03% | 17.51% | 29.70% | 4.17% |
| SMSspamCollection | Enron | 1 | 63.46% | 30.88% | 52.76% | 54.41% | 43.72% | 11.61% |
| | | 2 | 49.29% | 43.31% | 53.68% | 54.14% | 52.37% | 3.85% |

**Table 5: Result of Cross-dataset Poisoning Attacks on BERT**

| Training Set | Testing Set | Epoch | FNR | FPR | Accuracy | Precision | F1 Score | Train Loss |
|---|---|---|---|---|---|---|---|---|
| LingSpam | Enron | 1 | 14.80% | 40.40% | 72.46% | 68.02% | 75.65% | 12.11% |
| | | 2 | 19.31% | 40.07% | 70.36% | 67.01% | 73.22% | 1.73% |
| Enron | LingSpam | 1 | 5.10% | 30.49% | 73.90% | 39.41% | 55.69 | 7.50 % |
| | | 2 | 15.31% | 13.43% | 86.24% | 56.85% | 68.03% | 2.05% |
| LingSpam | SMSspamCollection | 1 | 0.00% | 86.34% | 25.20% | 15.16% | 26.33% | 14.00% |
| | | 2 | 0.00% | 82.61% | 28.43% | 15.73% | 27.19% | 2.25% |
| SMSspamCollection | LingSpam | 1 | 50.00% | 47.76% | 51.85% | 17.95% | 26.42% | 8.16% |
| | | 2 | 27.55% | 81.66% | 27.69% | 15.64% | 25.72% | 3.25% |
| Enron | SMSspamCollection | 1 | 0.67% | 79.92% | 30.67% | 16.09% | 27.69% | 7.96% |
| | | 2 | 0.00% | 79.92% | 30.76% | 16.18% | 27.85% | 2.07% |
| SMSspamCollection | Enron | 1 | 71.01% | 30.91% | 48.96% | 48.61% | 36.32% | 12.11% |
| | | 2 | 57.17% | 36.58% | 53.08% | 54.14% | 47.83% | 4.14% |

- *LLM Model Comparison:* It seems that GPT2 shows relatively lower FPR although suffers in FNR. So, it generally flags fewer messages as spam. It has a slightly better precision in detection. Such a spam filter has a "conservative" strategy in filtering messages. BERT is the opposite with higher FPR and lower FNR, showing a more "aggressive" style to more likely flag a message as spam. Such differences may not be conclusive with just three datasets. But this may shed light on the different styles of LLMs, useful for model selections that consider the system requirements influenced by FPR, FNR, and other criteria.
- *Dataset Comparison:* The challenge is obvious when the SMSspamCollection dataset is involved because of its difference from the two spam email datasets. The FNR rate is considerably higher when a model trained on SMSspamCollection is tested on an email dataset. Maybe this is because the strategies employed in SMS spams have evolved as a subset of those in spam emails. Therefore, training on the SMS spam messages does not provide the LLM models with adequate information to use in classifying emails. At the same time, the FNR performance of an LLM model trained on an email dataset does not suffer as much in classifying the messages in the SMSspamCollection dataset. However, the FPR result seems to deteriorate in all the experiments involving SMSspamCollection, no matter as the training or the testing set. An explanation may be that the legitimate emails and the legitimate short messages are quite different in their vocabulary and semantic style.

## 5.4 Additional Discussion

**Adversarial Attacks in Real-world Application Scenarios:** One of the most striking observations from the experiments is the significant impact of the injection position of the magic words or sentences on a model. When they are injected at the beginning of an email, they dominate the context used to influence how the information is processed in the downstream, distorting and overshadowing the real spam cues that the LLM model tries to understand.

In the real-world, this finding presents an interesting dilemma for attackers employing the "magic word" or "magic sentence" strategy. Placing such attacking material at the very beginning of the email can, as demonstrated in our experiments, effectively bypass spam filters. However, it comes with a significant trade-off: the higher visibility of such redundant information to human readers. Computer users very likely dismiss this email when the information up front appears nonsensical or irrelevant. On the contrary, burying the attacking material deeper within the email content makes such "spam cues" less visible to users but has diminishing effect to increase the chance of the spam email to slip through the filter. In the deployment of an LLM-based spam detection system, this insight can be useful to both the attacker and the defender in their working tactics.

**Generalization Challenges in Cross-Dataset Poisoning Attacks:** The experimentation underscores a fundamental limitation of LLM-based spam filters in that their dependency on the comprehensiveness of training data. Spam filters trained on insufficiently diverse datasets fail to generalize effectively to unseen or unfamiliar data. This is problematic in real-world scenarios where spam messages keep evolving and take various forms across different communication platforms. Importantly, we observed that models trained on longer, information-rich datasets like emails performed better when tested on shorter SMS messages, than the reverse. This suggests that models trained on more complex data are better equipped to adapt to "simpler" information processing tasks, while the opposite is not true.

## 6 Conclusion

In this study, we developed spam detection systems based on Large Language Models (LLMs) including GPT2 and BERT, evaluating their effectiveness across datasets including Enron, LingSpam, and SMSspamCollection. Our findings demonstrated that LLMs exhibit strong potential to capture nuanced patterns in textual data and leverage their semantic understanding to improve spam detection accuracy. However, their success depends on training on relevant and diverse datasets.

Facing adversarial attacks using specially selected "magic words" and "magic sentences", LLMs have displayed a degree of resilience when these distractions are added later in a spam message. However, their impact has proved that LLM-based spam filters are not immune from an attack strategy that purposefully modifies a message. Furthermore, cross-dataset testing has revealed the limitations of LLMs to adapt to differing linguistic structures and information distributions, suggesting that training with diverse datasets is critical to their practical deployment.

Future efforts should study newer LLM models with more representative datasets to understand their strength and limit. Additionally, experiments are needed to characterize the vulnerabilities of these models to a wide range of attacks and to evaluate defense mechanism to detect these attacks and mitigate their impact.